\documentclass[
superscriptaddress,
amsmath,amssymb,
aps,
twocolumn
]{revtex4-2}

\usepackage{color}
\usepackage[FIGTOPCAP]{subfigure}
\usepackage{graphicx}
\usepackage{dcolumn}
\usepackage{bm,braket}
\usepackage{comment}
\usepackage{here}
\usepackage{xcolor}  
\usepackage{float}

\newif\ifA

\begin {document}

\title{Density-Independent {\color{black}transient caging} in the high-density phase of motility-induced phase separation}

\author{Toranosuke Umemura}
\affiliation{%
	Department of Physics and Astronomy, Tokyo University of Science, Noda, Chiba 278-8510, Japan
}%

\author{Issei Sakai}
\affiliation{%
	Department of Physics and Astronomy, Tokyo University of Science, Noda, Chiba 278-8510, Japan
}%

\author{Takuma Akimoto}
\email{takuma@rs.tus.ac.jp}
\affiliation{%
	Department of Physics and Astronomy, Tokyo University of Science, Noda, Chiba 278-8510, Japan
}%



\date{\today}

\begin{abstract}
We investigate the nonequilibrium dynamics of active matter using a two-dimensional active Brownian particles model.
In these systems, self-propelled particles undergo motility-induced phase separation (MIPS), spontaneously segregating into dense and dilute phases.
We find that in the high-density phase, local particle mobility exhibits {\color{black}transient caging}, with diffusivity remaining unchanged despite variations in the global system density.
As global density increases further, the system undergoes a transition to a solid-like state through an intermediate regime with pronounced dynamical arrest. 
{\color{black}Our findings identify a distinct high-density regime characterized by transient caging and dynamical slowing down in a monodisperse active system, shedding new light on the connection between MIPS and nonequilibrium arrest.}
\end{abstract}

\maketitle


\section{Introduction}
Active matter consists of self-propelled particles or individual units that consume energy to produce movement or exert forces, leading to spontaneous motion and collective behavior \cite{TONER2005170,Lauga_2009,RevModPhys.85.1143,RevModPhys.88.045006,Gompper_2020}.
Unlike passive matter, which remains in thermodynamic equilibrium and moves only due to external forces or thermal fluctuations, active matter continuously consumes energy to drive self-propulsion and collective motion.
Examples include synthetic Janus particles \cite{walther2008janus,Vogel:2015aa}, motile bacterial cells \cite{doi:10.1073/pnas.0703530104}, and animal groups like bird flocks or fish schools \cite{Couzin:2005aa,doi:10.1073/pnas.1006874107,Sumpter:2008aa}, all of which display self-organization and collective dynamics.
These systems exhibit diverse collective phenomena, offering fundamental insights into nonequilibrium statistical mechanics.

One of the most striking behaviors in active matter is motility-induced phase separation (MIPS) \cite{annurev:/content/journals/10.1146/annurev-conmatphys-031214-014710}, where self-propelled particles segregate into dense and dilute phases due to the interplay between activity and interparticle interactions.
Unlike equilibrium phase separation, which typically requires attractive forces, MIPS occurs even in purely repulsive systems and has been observed in both simulations and experiments \cite{doi:10.1126/science.1230020,PhysRevLett.110.238301,doi:10.1098/rsta.2013.0372}.
Hydrodynamic interactions are known to influence MIPS, sometimes suppressing phase separation \cite{PhysRevE.90.032304}.
While MIPS has been widely studied, particularly regarding cluster formation and phase behavior, the dynamical properties of the high-density phase, especially in relation to glassy dynamics, remain an open question \cite{PhysRevResearch.7.013153}.

Diffusion in active matter systems is a key property that governs nonequilibrium transport and phase behavior.
The long-time diffusion coefficients in the active suspension decrease with increasing the density \cite{PhysRevLett.108.235702,PhysRevResearch.7.013153},
and at densities where MIPS occurs, they drop sharply \cite{PhysRevResearch.7.013153}. 
Within the high-density phase of MIPS, the mean squared displacement (MSD) is suppressed compared  to single-particle systems and exhibits subdiffusive, superdiffusive, and normal diffusive behaviors  at short, intermediate, and long times, respectively \cite{PhysRevLett.110.055701}.
While MIPS has been widely studied \cite{PhysRevResearch.7.013153}, the long-time diffusion properties of active particles in the high-density phase remain unclear.
{\color{black}In passive systems, caging by neighbors generally produces strong dynamical slowing down at high density, whereas in active systems self-propulsion adds a competing drive that may either enhance or suppress diffusion \cite{PhysRevLett.99.048102,PhysRevLett.108.235702}.
Quantifying diffusion in the dense MIPS phase is essential for distinguishing different high-density phases and for understanding nonequilibrium phase behavior.}

In two-dimensional active matter, distinguishing between glass-like dynamics and the hexatic phase is subtle but essential \cite{berthier2013non,bi2016motility,Miao2025}. The hexatic phase, characterized by quasi-long-range bond-orientational order and short-range translational order, remains fluid due to the absence of a finite shear modulus \cite{klamser2018thermodynamic, Digregorio2018}. In contrast, glassy behavior features structural disorder and dynamical arrest, often identified by a plateau in the MSD and suppressed particle mobility. 
{\color{black}Such nonequilibrium arrest has been explored mainly in polydisperse or frustrated active systems \cite{paoluzzi2024flocking,Keta2022}, whereas the dynamical constraints that arise inside the dense phase of monodisperse ABPs during MIPS coexistence remain less well characterized. Here we address this gap by quantifying local mobility in the dense MIPS phase, revealing transient caging and dynamical arrest signatures that precede full solidification even without geometric frustration.}

To this end, we employ a two-dimensional active Brownian particles (ABPs) model \cite{RevModPhys.85.1143, PhysRevLett.111.145702,Romanczuk:2012aa} to analyze the nonequilibrium dynamics across a range of global system densities, including those exhibiting MIPS.
We ask whether transport and local structural order in the high-density phase remain invariant or evolve as the global density increases, and how the system crosses over from MIPS to a solid-like arrested state. 
To characterize these behaviors, we use the bond-orientational order parameter and mean squared interparticle distance  (MSID), which quantify local structural order and particle motion in the high-density phase, respectively.
{\color{black}Our results demonstrate persistent transient caging within the MIPS dense phase and elucidate its smooth crossover to dynamical arrest at higher global densities, thereby shedding light on solidification in nonequilibrium active matter.}

\section{Model}

In the model, the system consists of $N$ self-propelled particles confined to a two-dimensional domain. Each particle's motion is governed by an overdamped Langevin equation
\begin{equation}
\gamma\dot{\bm{r}}_i(t) = \bm{\chi}_i(t) + \gamma s\bm{\omega}_i(t) - \nabla_i\sum_{j \neq i}\phi(|\bm{r}_{ij}|),
\end{equation}
where $ \gamma $ is the friction constant, $ \bm{r}_i(t)=(x_i(t),y_i(t))$ represents the position of particle $ i $ at time $ t $, $\bm{r}_{ij}=\bm{r}_i-\bm{r}_j$ represents the relative position vector between particles $i$ and $j$, and  $s$  is the constant speed of self-propulsion.
The term $ \bm{\chi}_i(t) $ represents thermal fluctuations described by two-dimensional white Gaussian noise, satisfying 
$
\left\langle \bm{\chi}_i(t) \right\rangle = 0 \quad \text{and} \quad \left\langle \bm{\chi}_i(t) \bm{\chi}_j(t') \right\rangle = 2k_{\rm B}T\gamma\delta(t-t')\bm{1}\delta_{ij}.
$
Here, $ k_{\rm B}$ denotes the Boltzmann constant, and $ T $ is the absolute temperature. The unit vector  $ \bm{\omega}_i(t) = (\cos\theta_i(t), \sin\theta_i(t)) $ determines the particle's orientation, which evolves according to
\begin{equation}
\tilde{\gamma}\dot{\theta}_i(t) = \xi_i,
\end{equation}
where $ \tilde{\gamma} $ is the rotational friction constant, and $ \xi_i $ is white Gaussian noise with 
$
\left\langle \xi_i(t) \right\rangle = 0 \quad \text{and} \quad \left\langle \xi_i(t) \xi_j(t') \right\rangle = 2k_{\rm B}T\tilde{\gamma}\delta(t-t')\delta_{ij}.
$
In the low-Reynolds number, the relation between $\gamma$ and $\tilde{\gamma}$ is $\tilde{\gamma}=\gamma\sigma^2/3$.
This noise represents the random reorientation of active particles and is independent of the translational thermal fluctuations. 

The interaction between particles is modeled using the Weeks-Chandler-Andersen (WCA) potential \cite{10.1063/1.1674820,PhysRevA.4.1597,PhysRevLett.126.208001,PhysRevE.103.052105,10.1063/1.3041421,PhysRevE.80.061101,Ben-Amotz:2004aa,10.1063/1.4923340}, a purely repulsive potential given by
\begin{equation}
	\phi(r_{ij}) =
	\begin{cases}
		4\varepsilon \left[ \left( \dfrac{\sigma}{r_{ij}} \right)^{12} - \left( \dfrac{\sigma}{r_{ij}} \right)^6 \right] + \varepsilon & (r_{ij} < 2^{1/6} \sigma) \\
		0 & \text{otherwise,}
	\end{cases}
\end{equation}
where  $\sigma$  represents the particle diameter, $\varepsilon$ is the interaction strength, and $r_{ij}=|\bm{r}_{ij}| $ denotes the distance between particle $i$ and $j$.
This potential ensures that particles experience only short-range repulsion, as the interaction is truncated at  $r_{ij} = 2^{1/6} \sigma$. Since the force derived from this potential vanishes beyond this cutoff, the interaction remains purely repulsive with no attractive component. 

To systematically characterize the system, we introduce two key dimensionless parameters.
The first is the P\'eclet number, defined as $\text{Pe} = 3s\tilde{\gamma}/(\sigma k_{\rm B}T)$, which quantifies the relative strength of self-propulsion compared to thermal fluctuations of the rotational diffusion.
A high P\'eclet number indicates that the system is dominated by self-propulsion rather than Brownian diffusion.
The second important parameter is the global number density, given by  $\rho = N / (L_x L_y)$, which represents the overall packing fraction of particles within the simulation domain.
Here, $L_x$ and $L_y$ are the system sizes.
These parameters allow us to explore the effects of activity and density on the system's phase behavior.

In our simulations, we set the temperature as $k_{\rm B} T = 1.0$, the particle diameter as $\sigma = 1.0$, and the P\'eclet number as $\text{Pe} = 120$, 
with the self-propulsion speed normalized to $s = 1.0$. This relatively high P\'eclet number ensures that self-propulsion dominates over thermal fluctuations, 
facilitating the exploration of MIPS \cite{PhysRevE.100.052604} and enabling us to investigate the potential emergence of {\color{black}transient caging} in the high-density phase.
The energy scale is chosen such that  $k_{\rm B} T/\varepsilon = 0.5$, ensuring a balance between thermal fluctuations and interaction strength.
The system consists of $N$ ABPs confined in a rectangular domain of size  $L_x = 120$  and  $L_y = 24$, with periodic boundary conditions.
This aspect ratio stabilizes the shape of dense clusters formed through MIPS and suppresses large fluctuations in their center-of-mass motion, enabling a clearer analysis of local particle dynamics.

\section{Results}

\begin{figure*}
	\centering
	\includegraphics[width=1.0\linewidth]{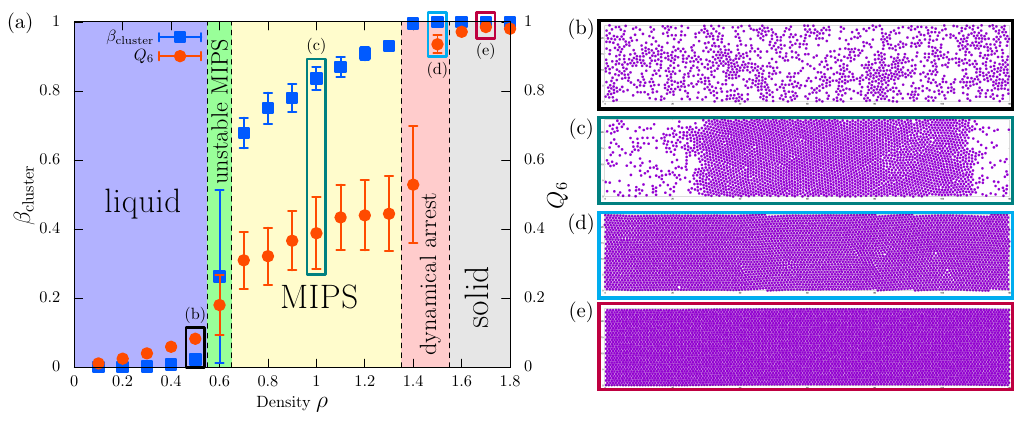}
	\caption{(a) Density dependence of the cluster fraction $\beta_{\text{cluster}}$ and bond-orientational order parameter  $ Q_6 $. Squares and circles show the results of numerical simulations for $\beta_{\text{cluster}}$ and $ Q_6 $, respectively. 
	Error bars represent the standard deviation. The background color indicates the system state: purple (liquid), green (unstable MIPS), yellow (MIPS), light red ({\color{black}dynamical arrest}), and gray (solid-like). (b)-(e) Representative snapshots at different global densities:$\rho = 0.5$, $1.0$, $1.5$, and $1.7$.}
	\label{fig: cluster_fraction_and_Q6}
\end{figure*}

\begin{figure}
	\centering
	\includegraphics[width=1.0\linewidth]{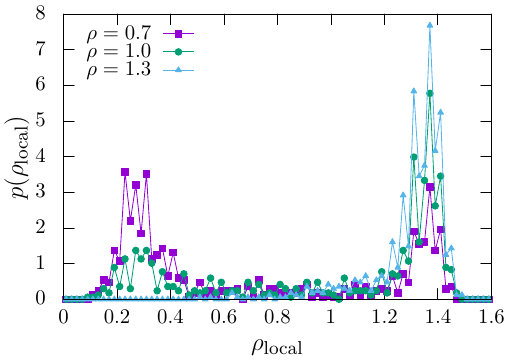}
	\caption{Probability distribution of the local density, $p(\rho_{\mathrm{local}})$, measured in the steady state for three global densities $\rho=0.7,\,1.0$, and 1.3. The distributions are bimodal, with a low-density peak (gas phase) and a high-density peak (dense phase), demonstrating phase coexistence in the MIPS regime.  }
	\label{fig: pdf local density}
\end{figure}

\subsection{Cluster analysis and phase stability in MIPS}
We analyze the structural and dynamical properties of the high-density phase in MIPS by systematically varying the global system density.
To characterize the onset of clustering, we compute the fraction of particles within clusters, $\beta_{\text{cluster}}$, defined as $\beta_{\text{cluster}}  = \left\langle N_{\text{cluster}}\right\rangle /N,$ where $ N_{\text{cluster}} $ is the number of particles within clusters.
A particle is considered to belong to a cluster if it has at least four neighboring particles within a radius $\sigma $. 

Figure~\ref{fig: cluster_fraction_and_Q6} illustrates the variation of $\beta_{\text{cluster}} $ with global density.
At low densities ($\rho<0.6$), $\beta_{\text{cluster}} \approx 0 $, indicating no significant clustering.
In this phase, the system remains a homogeneous liquid, where nearly all particles move independently.
At $\rho=0.6$, $\beta_{\text{cluster}}$ exhibits large fluctuations because MIPS repeatedly emerges and disappears.
This indicates that the high density phase in MIPS is unstable.
As the density increases beyond $\rho=0.7$, $ \beta_{\text{cluster}}$ exhibits a sharp increase and only slight fluctuations.
These fluctuations are due to the exchange of particles between the high-density and low-density phases.
The error bars are smaller than the one at $\rho=0.6$, suggesting that the high-density phase in MIPS is stable. 
For $ \rho \geq 1.4$, nearly all particles belong to clusters ($\beta_{\text{cluster}} \approx 1 $), signifying a transition to a solid-like phase where particle mobility is highly constrained.

{\color{black}Figure~\ref{fig: pdf local density} shows the probability distribution of the local density. Within the MIPS regime, increasing the global density $\rho$ primarily increases the weight of the high-density peak in $p(\rho_{\mathrm{local}})$, while the peak position itself remains nearly unchanged. This indicates that added particles are accommodated mainly by expanding the area occupied by the dense phase, rather than by further compressing it. Hence, the dense-phase local density $\rho_{\mathrm{HD}}$ varies only weakly with $\rho$, consistent with gas-dense coexistence at roughly constant packing.}

\subsection{Structural order analysis via $Q_6$ parameter}
To explore the  structural properties of the system, we analyze structural ordering using the bond-orientational order parameter $Q_6$ \cite{PhysRevLett.108.168301,PhysRevLett.112.118101}, a widely used metric that quantifies the degree of local hexagonal ordering among neighboring particles. This parameter is particularly sensitive to six-fold symmetry and is commonly employed to detect crystalline and hexatic structures in two-dimensional systems.
This parameter is computed based on the relative angles between neighboring particles and is defined as
\begin{equation}
	Q_6 = \frac{1}{6N}\left\langle\sum_{i=1}^N\sum_{j \neq i, \, r_{ij} < \sigma} \cos^2(6\alpha_{ij}) \right\rangle.
	\label{eq: Q6}
\end{equation}
Here,  $\alpha_{ij} $ represents the relative angle between particles $i$ and $j$.
The neighboring particles are defined as those for which the particle-pair distance $r_{ij}$ is smaller than $ \sigma $.
The order parameter $Q_6$ quantifies the extent of hexagonal order in the system.
For a perfect hexagonal arrangement, $Q_6$ approaches 1, indicating a highly ordered structure. Conversely, for a disordered state, $Q_6$ is close to 0, reflecting the absence of regular particle alignment.
This parameter has been widely used in studies of order in liquids and glasses \cite{PhysRevB.28.784}.

Figure~\ref{fig: cluster_fraction_and_Q6} shows how the order parameter $Q_6$ varies with the global density. At low densities ($\rho < 0.6$), $Q_6$ takes small values, indicating that the system lacks overall structural regularity---consistent with a liquid state. 
As the density increases into the intermediate-density range ($0.6 \le \rho < 1.4$),  $Q_6$  gradually increases, suggesting the emergence of short-range order.
In this regime, local particle arrangements become more regular, but global stractural order is not achieved.
Moreover, the wide error bars indicate large fluctuations in particle arrangement and the cluster size, suggesting that the system remains structurally disordered at larger scales. 
As shown in the cluster fraction analysis, MIPS occurs in this density range, leading to the coexistence of a high-density phase and a low-density phase.
At higher densities ($ \rho = 1.4$ and $1.5$), particles {\color{black}are densely packed and develop strong local sixfold order, but the remaining spread in $Q_6$ indicates that this order is not yet fully stabilized across the system.}
As the density increases beyond $\rho = 1.5$, $Q_6$ approaches unity, and the reduced spread of error bars suggests that the order becomes more stable.
Therefore, when $\rho \geq 1.4$, all particles belong to clusters. However, short-range order is unstable for $1.4 \leq \rho < 1.6$, while it becomes stable for $\rho \geq 1.6$.


\subsection{ Probing dynamical arrest using the MSID}

{\color{black}Long-time diffusivity is a sensitive indicator of dynamical slowing in dense systems. We therefore quantify particle mobility and structural constraints in the high-density phase of MIPS.}
A common method for evaluating diffusivity is to analyze the MSD of a single particle. 
In the active suspension, the MSD of the entire system and the MSD in the high-density phase of MIPS have been numerically investigated \cite{PhysRevResearch.7.013153,PhysRevLett.110.055701}.
However, in the high-density phase, the entire cluster undergoes diffusion, leading to collective motion and system-wide fluctuations.
As a result, the MSD of an individual particle reflects both its intrinsic motion and the center-of-mass motion of the entire system, making it difficult to isolate true particle diffusivity.
This issue is analogous to that observed in small systems such as lipid bilayers, where global fluctuations obscure individual particle motion, complicating diffusivity measurements \cite{PhysRevLett.107.178103}.
To overcome this limitation, we analyze the MSID, which provides a more reliable measure of the relative motion of particle pairs and better captures {\color{black}mobility} in dense phases.
The MSID is defined as
\begin{equation}
	\langle \Delta d^2(t) \rangle = \left\langle \frac{1}{N_{\text{pair}}} \sum_{i<j} \left| \bm{r}_{ij}(t) - \bm{r}_{ij}(0) \right|^2 \right\rangle
	\label{eq: MSID}
\end{equation}
where $N_{\text{pair}}$ is the total number of particle pairs that  remained continuously in the high-density phase for the entire observation period.
By analyzing MSID, {\color{black}we directly track how particle separations evolve over time, providing a sensitive measure of dynamical constraints and transient caging in the dense phase.}
Notably, the MSID exhibits a plateau when cage effects confine particles locally, even if the center-of-mass motion of the cluster is not restricted. This distinguishes the relative motion of particles from global drift and highlights the emergence of {\color{black}dynamical arrest} within the dense phase.
In Ref.~\cite{PhysRevLett.108.168301}, the MSID was employed to determine whether the active suspension exhibits solid-like or liquid-like behavior.

\subsubsection{ Connecting the MSID and MSD under harmonic confinement}
In the high-density phase, particles are restricted by their neighbors. In the short-time regime, where particles do not escape their cages, they effectively move independently within their local environments. We approximate their motions as being constrained by independent harmonic potentials and assume that their motions are uncorrelated. Consequently, the MSID can be derived from the MSD via  $\langle \Delta d^2(t) \rangle = 2 \langle \Delta {\bm r}^2(t) \rangle$, where $\Delta {\bm r}(t)= {\bm r}(t) - {\bm r}(0)$ is the displacement vector. 

When the initial positions of the particles are sampled from the steady state, the MSD of an ABP in a harmonic potential is given by
\cite{PhysRevLett.129.158001,halder2025interplaytimescalesgovernsresidual}
	\begin{equation}
		\begin{split}
			\Braket{\Delta {\bm r}^2(t)}
			=&\frac{4k_{\rm B}T}{\kappa(\rho_{\text{HD}})}\left[1-e^{-t/\tau_k}\right]\\
			&+\frac{2s^2\tau_k^2\tau_R^2}{\tau_R+\tau_k}\left[1-\frac{\tau_Re^{-t/\tau_R}-\tau_ke^{-t/\tau_k}}{\tau_R-\tau_k}\right],
		\end{split}
		\label{MSD in the harmonic potential}
	\end{equation}
where $\kappa(\rho_{\text{HD}})$ is the effective spring constant used to approximate the confining effects of neighboring particles, parameterized by the local density $\rho_{\text{HD}}$ in the high-density phase (See Appendix~A). Here, $\tau_k=\gamma/\kappa(\rho_{\text{HD}})$ is the characteristic relaxation time for the particle's motion in the harmonic potential, and $\tau_R=\tilde{\gamma}/k_{\rm B}T$ is the rotational relaxation time of the particle, with $\tau_k \ll \tau_R$.
The MSD initially grows linearly with time at short times ($t \ll \tau_k$), consistent with free diffusion. At intermediate times ($t \sim \tau_k$), the MSD reaches its first plateau, corresponding to the particle being trapped by the harmonic potential due to thermal fluctuations. At longer times ($\tau_k \ll t \ll \tau_R$), the MSD remains at this plateau as the particle's orientation remains relatively constant. Finally, at times comparable to or greater than $\tau_R$, the rotational diffusion of the propulsion direction allows the particle to explore new configurations, leading to an additional contribution to the MSD from self-propulsion. This contribution creates a second plateau associated with the rotational fluctuations of the particle's propulsion direction.


\begin{figure}
	\centering
	\includegraphics[width=1.0\linewidth]{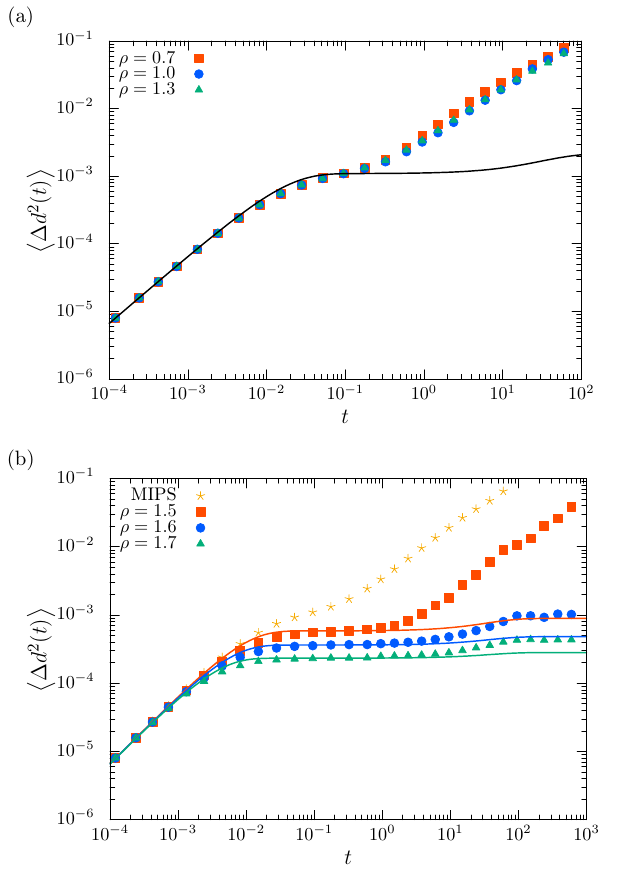}
	\caption{(a) Mean squared interparticle distance (MSID) in the high-density phase of MIPS for different global density ($\rho=0.7, 1.0,$ and $1.3$). Squares, circles, and triangles 
	correspond to the numerical simulation results for 
	   $\rho=0.7, 1.0,$ and $1.3$, respectively. 
	The solid line represents Eq.~\eqref{MSD in the harmonic potential} with $\rho_{\text{HD}}=1.38$, the density of the high-density phase obtained from the numerical simulations.
	(b) MSID at higher global densities ($\rho=1.5$, $1.6$, and $1.7$). Squares, circles, and triangles correspond to the numerical simulation results for $\rho=1.5$, $1.6$, and $1.7$, respectively. 
	Crosses show the numerical simulation result for the high-density phase at $\rho=1.3$ as a reference.
	The solid lines represent Eq.~\eqref{MSD in the harmonic potential}.}
	\label{fig:msd_analysis}
\end{figure}

\subsubsection{MSID behavior in the MIPS regime $(0.7 \leq \rho < 1.4)$}

In the MIPS regime, we analyze the MSID for particles belonging to the dense phase of the MIPS coexistence. At short times ($t\ll\tau_k$), the MSID increases linearly with time, consistent with free diffusion where interparticle interactions have minimal influence.
The relationship for short times is given by
\begin{equation}
	\langle \Delta d^2(t) \rangle \sim \frac{8k_{\rm B} T}{\gamma}t,
	\label{short_theory}
\end{equation}
which follows from the fact that for two-dimensional Brownian motion, the MSD of a single particle is $ 4 k_{\rm B} T t/\gamma$. 
Since MSID measures the displacement between two particles, rather than an individual one, the factor is doubled, leading to Eq.~\eqref{short_theory}.

At intermediate times ($\tau_k\ll t\ll\tau_R$), interparticle interactions constrain particle motion, leading to the emergence of a plateau in $\langle \Delta d^2(t) \rangle$ 
{\color{black}that signals transient caging in the dense phase} \cite{PhysRevE.58.3515,Levashov_2021}.
For $t/\tau_k\gg 1$ and $t/\tau_R\ll 1$, Eq.~\eqref{MSD in the harmonic potential} is given by $\langle \Delta {\bm r}^2(t) \rangle \cong 4k_{\rm B} T/\kappa(\rho_{\text{HD}})$.
Thus, the MSID at the plateau is expressed as
\begin{equation}
	\langle \Delta d^2(t) \rangle \cong \frac{8 k_{\rm B} T}{\kappa(\rho_{\text{HD}})}.
	\label{MSID_plateau}
\end{equation}
This relation coincides with the long-time limit of the MSID for Brownian particles in a harmonic potential.
The agreement arises because the system is observed prior to the onset of rotational relaxation, where the contribution from ballistic motion remains negligible. 
{\color{black}The MSID plateau is relatively short and does not persist over a wide temporal range, distinguishing it from the long-lived caging typical of glassy systems. Nonetheless, its presence provides clear evidence of transient caging in the high-density phase. Notably, the plateau height and duration remain essentially invariant as the global density is increased within the MIPS regime, and increase sharply only after the dense phase becomes homogeneous at higher global densities, signalling the crossover to dynamical arrest.}

At long times, particles escape their cages \cite{doi:10.1098/rsif.2013.0726,PhysRevLett.112.220602,C3SM52469H}, resulting in an increase in the MSID.
This behavior reflects the onset of fluidity within the high-density phase, where particles gradually regain mobility and transition back to diffusive motion.
This transition---from free diffusion at short times to constrained motion at intermediate times, and finally back to diffusion at long times---{\color{black}indicates transient caging rather than a fully arrested glassy state.}

In MIPS, the system segregates into coexisting dense and dilute phases, and the density of the high-density phase $\rho_{\text{HD}}$ typically remains relatively stable, with only weak dependence on the global system density $\rho$ \cite{PhysRevE.100.052604,PhysRevLett.110.055701}.
Consequently, the MSID plateau remains constant within the high-density phase, regardless of variations in the global system density.
As shown in Fig.~\ref{fig:msd_analysis}(a), in the range  $0.7 \leq \rho < 1.4$  where MIPS occurs, the MSID remains independent of the global system density $\rho $.
This finding suggests that diffusivity in the high-density phase remains unchanged, even as the overall system density increases.

\subsection{ {\color{black}Dynamical arrest} in the homogeneous high-density regime $(1.4 \leq \rho <1.6)$}

At higher global densities, the high-density phase spans the entire system, forming a homogeneous phase. In this regime, the MSID exhibits a pronounced plateau at intermediate times [see Fig.~\ref{fig:msd_analysis}(b)], indicative of constrained particle motion due to caging. At $\rho = 1.5$, a clear plateau emerges in the MSID, and its duration becomes progressively longer with increasing density.
Following this plateau, the MSID resumes linear growth, suggesting that particles eventually escape their cages and regain mobility. The temporal evolution---initial growth, extended plateau, and diffusive recovery---signals the presence of {\color{black}dynamical arrest} in the density range $1.4 \leq \rho < 1.6$.
Analysis of the intermediate scattering function reveals that the structural relaxation time becomes progressively prolonged with increasing density, {\color{black}indicating pronounced dynamical slowing down in the dense phase} (see Appendix B).
During this regime, the bond-orientational order parameter $Q_6$ approaches unity, reflecting strong local sixfold symmetry. 
{\color{black}While this trend is consistent with increasing structural organization in the dense phase, the present analysis does not allow us to determine the exact nature of the ordered state (e.g., hexatic versus solid). In parallel, the MSID shows a plateau and a pronounced suppression of diffusivity, indicating the emergence of caging and strongly constrained dynamics. Taken together, these observations show that local order can build up alongside substantial dynamical constraints, which motivates analyzing dynamic (e.g., MSID, self-intermediate scattering function) and static (e.g., $Q_6$) observables together when discussing high-density arrested or solidifying regimes in active matter.
}

\subsubsection{ Solidification in the Homogeneous High-Density Regime $(\rho \geq 1.6)$}

In contrast, for $\rho = 1.6$ and $1.7$, the MSID exhibits a slight increase followed by the emergence of a second plateau, suggesting a transition into a solid-like phase.
This behavior is qualitatively similar to the MSD of an ABP confined in a harmonic potential.
In this density regime ($\rho \geq 1.6$), particles are confined by their neighbors, indicating that the system is in a solid-like state.
However, the value of the second plateau is larger than that observed for an ABP in a static harmonic potential.
In this study, we employed a mean-field approximation by modeling the particle as an ABP confined in a static harmonic potential.
In contrast, in the actual system, the confinement can be regarded as a dynamic harmonic potential, where the minimum position fluctuates diffusively.
These effects lead to an enhancement of the second plateau in the MSID compared to the static harmonic potential case.
In this regime, the bond-orientational order parameter $Q_6$ is almost unity, indicating that particles are arranged with strong local sixfold symmetry, 
 characteristic of a solid-like phase.
{\color{black}Notably, our system shows a clear plateau in the MSID together with suppressed diffusivity. These observations are consistent with a crossover from transient caging in the dense MIPS phase to solidification through dynamical arrest as the global density is increased.
}

\section{Conclusion}

In this study, we analyzed the dynamical behavior of two-dimensional active Brownian particles as the global density increases, and observed a progression from a liquid-like homogeneous state to MIPS and, at still higher densities, to a solid-like arrested regime. 
Using the bond-orientational order parameter $Q_6$ and the MSID, we characterized the concurrent evolution of local structural order and particle motion in the dense phase. A central finding is that the high-density branch of MIPS exhibits robust transient caging: particles display constrained mobility and an MSID plateau while long-range order is not established. Importantly, within the MIPS coexistence regime the plateau height and duration remain essentially independent of the global density, indicating that transport properties inside the dense phase are largely insensitive to changes in the overall density. Only after the dense phase becomes spatially homogeneous at higher global densities do the caging times grow rapidly, signaling a crossover to dynamical arrest.

These results clarify how transient caging in the MIPS dense phase connects to solidification in nonequilibrium active matter, and highlight the value of jointly analyzing dynamical and structural observables when mapping high-density active regimes. Beyond the present ABP model, the same approach can be applied to more complex active systems---including biological assemblies and synthetic self-propelled particles---providing a general framework for studying nonequilibrium phase behavior and dynamical crossovers in dense active matter.

\appendix

\section{Effective interparticle interactions}

In our analysis, we approximate the effective interparticle interactions in the high-density phase using a mean-field harmonic potential. The mean-field approach represents the averaged effect of neighboring particles on a given particle. Specifically, this interaction is modeled by a harmonic potential with a spring constant $ \kappa $, which quantifies the strength of the restoring force that constrains particles near their equilibrium positions. To determine $ \kappa $, we compare it to the Weeks-Chandler-Andersen (WCA) potential, which describes the short-range repulsive interactions in our system.

\begin{figure}[htbp]
	\centering
	\includegraphics[width=0.3\linewidth]{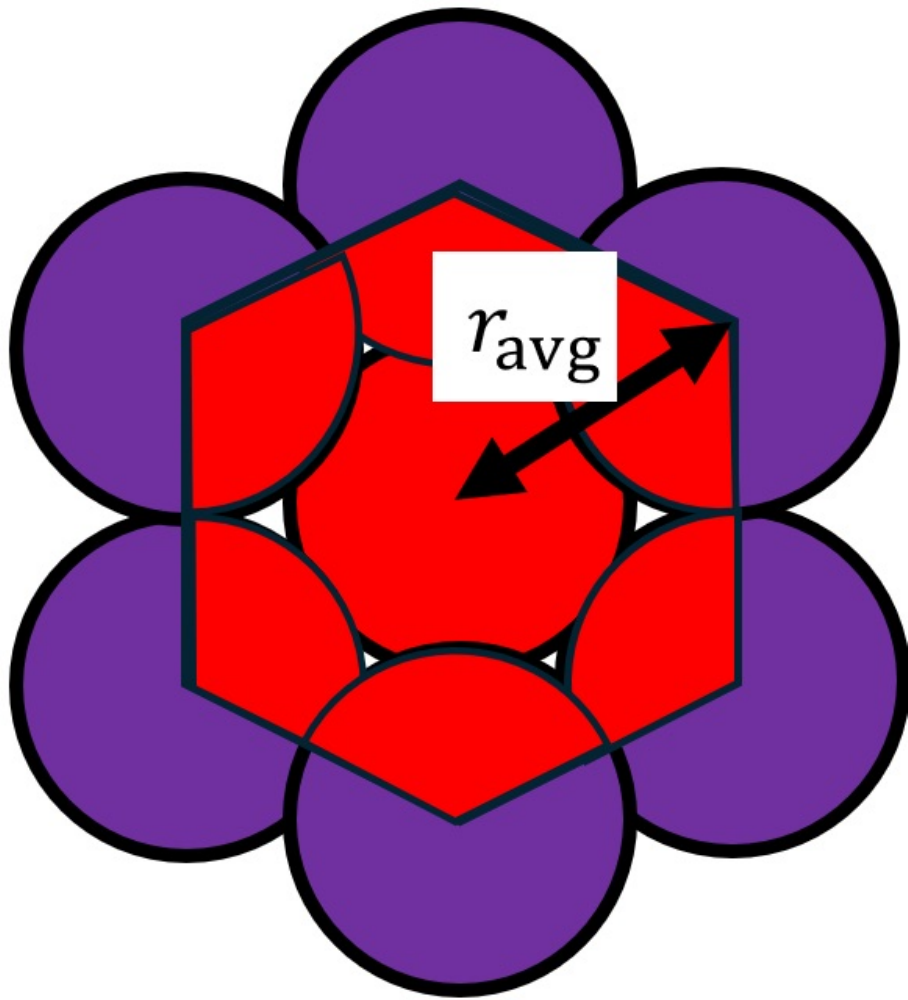}
	\caption{Hexagonal close-packed structure.
	$ r_{\text{avg}} $ is the optimal particle distance in the hexagonal close-packing structure.}
	\label{fig:hexagon}
\end{figure}

The WCA potential is a truncated and shifted form of the Lennard-Jones potential. 
This potential effectively prevents overlapping of particles while allowing for local fluctuations around the equilibrium separation distance  $r_{\text{avg}} $ (see Fig.~\ref{fig:hexagon}). To approximate the spring constant  $\kappa $, we expand the WCA potential around the equilibrium distance and take the second derivative:
\begin{equation}
\kappa = \left. \frac{d^2 \phi_{\text{WCA}}}{dr^2} \right|_{r = r_{\text{avg}}}.
\end{equation}
The spring constant is determined by comparing it to the  WCA potential and is given by
\begin{widetext}
\begin{equation}
	\begin{split}
		\kappa
		=\left[\phi''(r_1+r_{\text{avg}})+\phi''(r_2-r_{\text{avg}})\right]_{r_1=0,r_2=0}
		=1248\varepsilon\frac{\sigma^{12}}{r_{\text{avg}}^{14}}-336\varepsilon\frac{\sigma^{6}}{r_{\text{avg}}^{8}}
	\end{split}
\end{equation}
\end{widetext}
Here, $ r_{\text{avg}} $ is the optimal particle distance in the hexagonal close-packed (HCP) structure, which depends on the density $\rho$. 

\begin{figure}[htbp]
	\centering
	\includegraphics[width=1.0\linewidth]{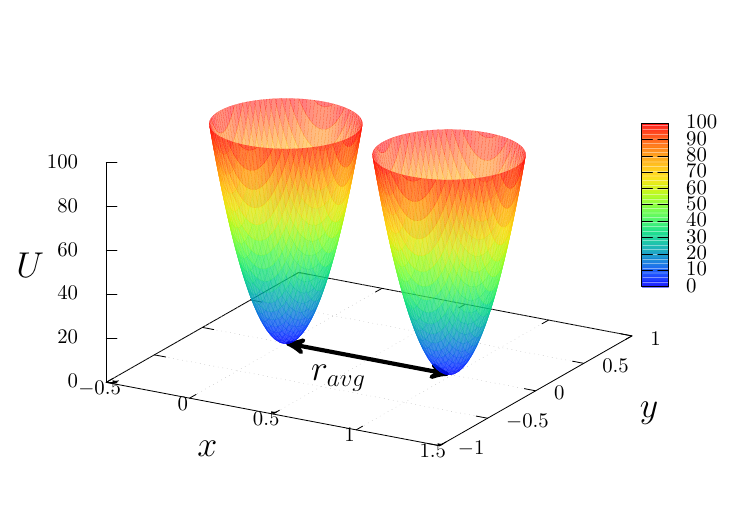}
	\caption{Two independent harmonic potentials. The distance between the two potential minima is $ r_{\text{avg}} $.
	By approximating the motion of ABPs in the high-density phase to the motion of particles in this potential, the plateau value of the MSID can be calculated.}
	\label{fig:potential_tikz}
\end{figure}

Consider two independent Brownian particles under harmonic potentials [see Fig.~\ref{fig:potential_tikz}]. The equilibrium distance  $r_{\text{avg}} $ in a HCP structure is determined by the density in the high-density phase $\rho_{\text{HD}}$ [see Fig.~\ref{fig:hexagon}]. In an HCP arrangement, it follows the relation
\begin{equation}
	\rho_{\text{HD}} = \frac{3}{\frac{6\sqrt{3}}{4}r_{\text{avg}}^2}.
\end{equation}
Rearranging this expression, we obtain $r_{\text{avg}} = \sqrt{{2}/{\sqrt{3} \rho_{\text{HD}}}}$. 

\section{Self-intermediate scattering function}

\begin{figure}[h]
	\centering
	\includegraphics[width=1.0\linewidth]{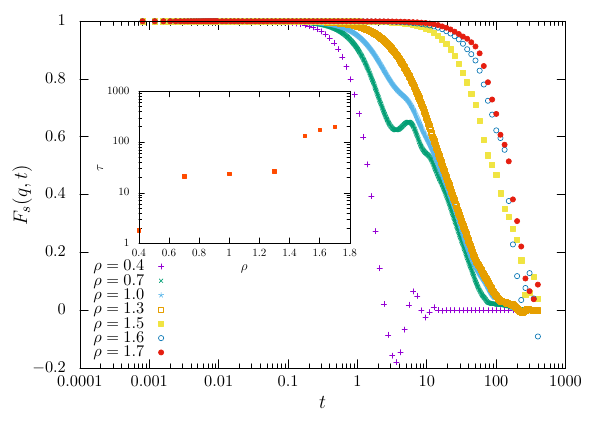}
	\caption{Self-intermediate scattering function. Symbols represent the numerical simulation results. The inset shows the relaxation time as a function of the particle density.}
	\label{fig:ISF}
\end{figure}

Here, we analyze the self-intermediate scattering function,
\begin{equation}
F_s(q,t) = \frac{1}{N} \sum_{j=1}^{L} \left\langle e^{i \mathbf{q} \cdot [\mathbf{r}_j(t) - \mathbf{r}_j(0)]} \right\rangle,
\end{equation}
where $\mathbf{q}$ is the wave vector.
The dynamics were evaluated for the entire system using $q = |\mathbf{q}|$ values close to the first peak of the static structure factor.
Figure~\ref{fig:ISF} shows the numerical results of the self-intermediate scattering function.
The relaxation time $\tau$ systematically increases with increasing particle density.
In the MIPS regime, a non-monotonic behavior appears, which we attribute to contributions from the coexisting dilute phase.
Because our analysis averages over the entire system, we attribute this to the liquid phase; its influence decreases as the density increases.
The intermediate scattering function exhibits a subtle plateau close to unity, corresponding to the MSID plateau within the same temporal regime. 
Notably, the plateau value deviates from that expected in a purely fluid state, suggesting that it reflects a genuine caging effect. The final decay at long times is likely due to the diffusive motion of the center of mass of the system.
Because the self-intermediate scattering function did not reveal significant structural distinctions across densities, we instead used the MSID to characterize local structure and dynamic arrest more effectively.

\begin{acknowledgments}
T.A. was supported by JSPS Grant-in-Aid for Scientific Research (No. C JP21K033920).
\end{acknowledgments}


%

\end{document}